\documentclass[aps,prl,twocolumn,showpacs,preprintnumbers,amsmath,amssymb,superscriptaddress]{revtex4}

\usepackage{graphicx}
\usepackage{epsfig}
\usepackage{amsmath}
\usepackage{amssymb}
\usepackage{textcomp}
\usepackage{dcolumn}
\usepackage{bm}
\usepackage{braket}
\usepackage[usenames,dvipsnames]{xcolor}
\usepackage[colorlinks=true,citecolor=MidnightBlue,linkcolor=MidnightBlue,urlcolor=MidnightBlue]{hyperref}
\usepackage{siunitx}

\PassOptionsToPackage{numbers,sort&compress}{natbib}

\makeatletter
\g@addto@macro\normalsize{%
  \setlength\abovedisplayskip{4pt}
  \setlength\belowdisplayskip{4pt}
  \setlength\abovedisplayshortskip{4pt}
  \setlength\belowdisplayshortskip{4pt}
}
\makeatother

\newcommand{\BibitemShut}[1]{} 

\begin{document}

\newcommand*{\MAINZ}{QUANTUM, Institut f\"ur Physik, Johannes Gutenberg-Universit\"at Mainz, Staudingerweg 7, 55128 Mainz, Germany}
\newcommand*{\ERLANGEN}{Institut f\"ur Optik, Information und Photonik, Friedrich-Alexander Universit\"at Erlangen-N\"urnberg, Staudtstr. 1, 91058 Erlangen, Germany}
\newcommand*{\SAOT}{Erlangen Graduate School in Advanced Optical Technologies (SAOT), Friedrich-Alexander Universit\"at Erlangen-N\"urnberg, Paul-Gordan-Str. 6, 91052 Erlangen, Germany}
\homepage{http://www.quantenbit.de}

\title{Imaging trapped ion structures via fluorescence cross-correlation detection}
\author{Stefan Richter}\email{stefan.michael.richter@fau.de}\affiliation{\ERLANGEN}\affiliation{\SAOT}
\author{Sebastian Wolf}\affiliation{\MAINZ}
\author{Joachim von Zanthier}\affiliation{\ERLANGEN}\affiliation{\SAOT}
\author{Ferdinand Schmidt-Kaler}\affiliation{\MAINZ}

\date{\today}

\begin{abstract}
Cross-correlation signals are recorded from fluorescence photons scattered in free space off a trapped ion structure. The analysis of the signal allows for unambiguously revealing the spatial frequency, thus the distance, as well as the spatial alignment of the ions. For the case of two ions we obtain from the cross-correlations a spatial frequency $f_\text{spatial}=1490 \pm 2_{stat.}\pm 8_{syst.}\,\text{rad}^{-1}$, where the statistical uncertainty improves with the integrated number of correlation events as $N^{-0.51\pm0.06}$. We independently determine the spatial frequency to be $1494\pm 11\,\text{rad}^{-1}$, proving excellent agreement. Expanding our method to the case of three ions, we demonstrate its functionality for two-dimensional arrays of emitters of indistinguishable photons, serving as a model system to yield structural information where direct imaging techniques fail. 
\end{abstract}

\maketitle

Intensity correlations introduced by R. Hanbury Brown and R. Q. Twiss more than 60 years ago \cite{HBT177,HBT178}  have served for determining the angular diameter of individual stars or distances between stars \cite{Brown1968a,Brown1968b,Brown1974}. 
In combination with the concept of higher order photon coherences - developed by R. Glauber \cite{Glauber1963a,Glauber1963b} - these experiments paved the way for quantum optics \cite{Glauber2006}. Since then intensity or photon auto-correlation measurements have been employed for characterizing light sources \cite{Loudon2000,Antibunching2016}, e.g., thermal sources or single photon sources (SPE) such as single atoms, ions, color centers, molecules or quantum dots. 
Cross-correlations of fluorescence photons emanating from independent SPEs have also been measured, for demonstrating the Hong-Ou-Mandel effect \cite{HOM1987} via two-photon interference \cite{kaltenbaek2006,beugnon2006,maunz2007,yamamoto2009,sandoghdar2010,solomon2010,hanson2012}, or for producing remote entanglement of emitters via projective measurements of photons \cite{Monroe2007,Weinfurter2012,Hanson2013,Blatt2013,Imamoglu2016,Atatuere2017}. 
Yet, in all of these cases single spatial modes have been picked out for collecting the photons. 
This approach, however, inhibits the observation of a genuine spatial interference pattern based on second order coherence that would reveal the information about the SPE arrangement.
Consequently, photon cross-correlations from microscopic SPE structures have not been recorded so far for obtaining spatial information about the emitter distribution. 

\begin{figure}
\includegraphics[width=0.45\textwidth]{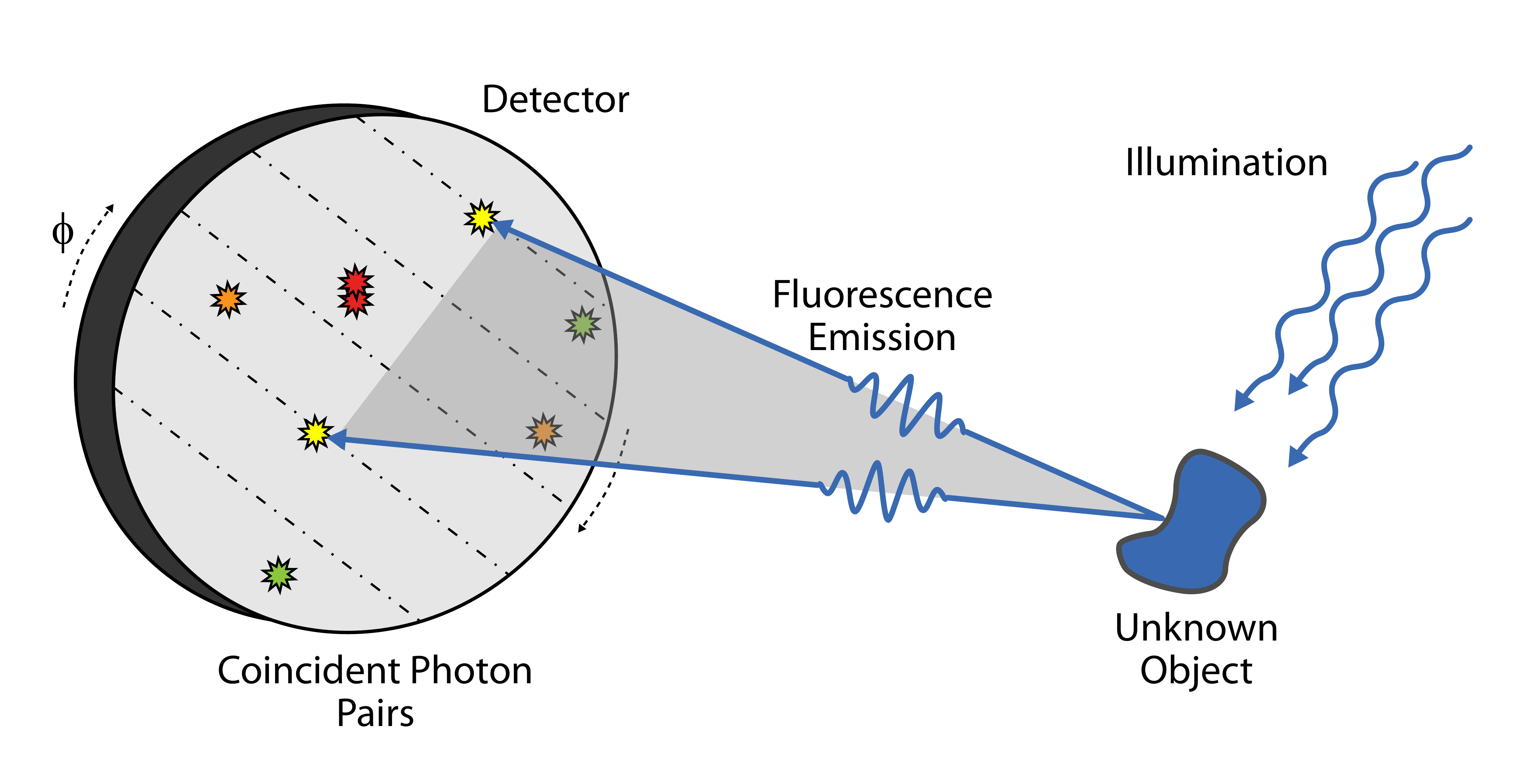}
\caption{The spatial arrangement of SPEs in a fluorescing object (blue) can be determined by measuring the spatial photon cross-correlations.
Coincident two-photon events (pairs of stars of equal color) are recorded by an ultra-fast camera in the far field. For low coincident rates, the 2D distribution of the recorded data can be binned along a line (dashed-dotted) to a 1D distribution. The line of binning can be rotated by an angle $\phi$. 
\\ \\}
\label{Fig1}
\end{figure}

Here we report the measurement of cross-correlations using fluorescence photons emitted into free space. 
The data analysis of the two-photon interference pattern allows for fully extracting the spatial arrangement of the SPEs, thus the number of SPEs, their spatial frequencies and their absolute orientation in space. 
Demonstrated here with a model system of a trapped ion structure, our experiment may serve for elucidating far-field imaging techniques based on fluorescence photon cross-correlations. 
We anticipate the scheme to be relevant for X-ray structure analysis of complex molecules or clusters, when direct imaging techniques fail and lens-less observation of incoherently scattered photons is advantageous \cite{Schneider2018,Classen2017}. 
Here, if fluorescence light is scattered into a large solid angle, high momentum transfer vectors can be accessed, enabling potentially higher resolution as compared to commonly used coherent diffraction imaging techniques \cite{Classen2017}. 
Our newly demonstrated structure analysis method might also be adapted to nanooptics for resolving SPE arrays closer spaced than the diffraction limit \cite{thiel2007quantum,Oppel2012}. It may further serve for imaging situations in the life sciences when scattering in diffusive or turbulent media inhibits obtaining structural information about the source arrangement \cite{Katz2014,Tian2018}. In fact, overcoming the turbulences of the atmosphere was highlighted as a major advantage of two-photon interferometry when proposed for astronomical observations \cite{HBT177,HBT178,HBT218,HBT1968}. 

\begin{figure*}
\includegraphics[width=\textwidth]{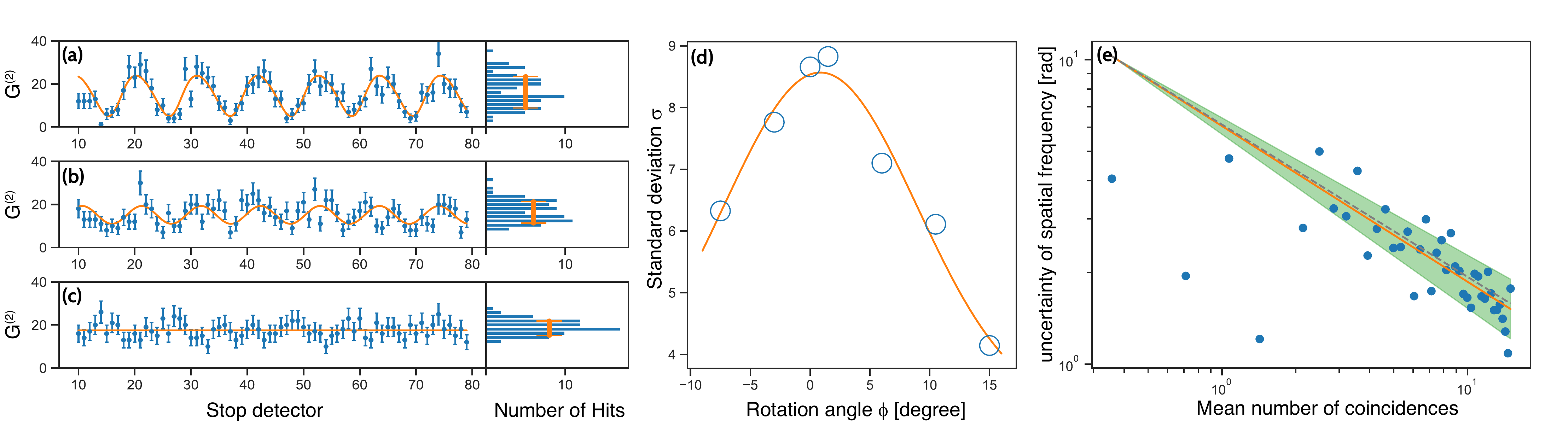}
\caption{ (a)-(c) One-dimensional cross-correlation signal and corresponding cosine-fit for different rotation angels $\phi=0^{\circ}$~(a), $\phi=10.5^{\circ}$~(b), and $\phi=15^{\circ}$~(c); plotting the signal in a histogram yields different standard deviations. Maximizing the standard deviation allows to find the optimal rotation angle (for details see text).
(d) Standard deviation as a function of rotation angle $\phi$. The maximum at $\phi = 0.86^{\circ} \pm 0.31 ^{\circ}$ determines the absolute orientation of the structure. 
(e) Uncertainty of the spatial frequency $f_\text{spatial}$ as a function of the number of coincidences $N$; the fit (solid line) follows $N^{-0.51}$, and for comparison $N^{-0.5}$ (dashed line); the $1 \sigma$ uncertainty of the fit of 0.06 is indicated (shaded area)}
\label{Fig2}
\end{figure*}

\begin{figure}
\includegraphics[width=0.45\textwidth]{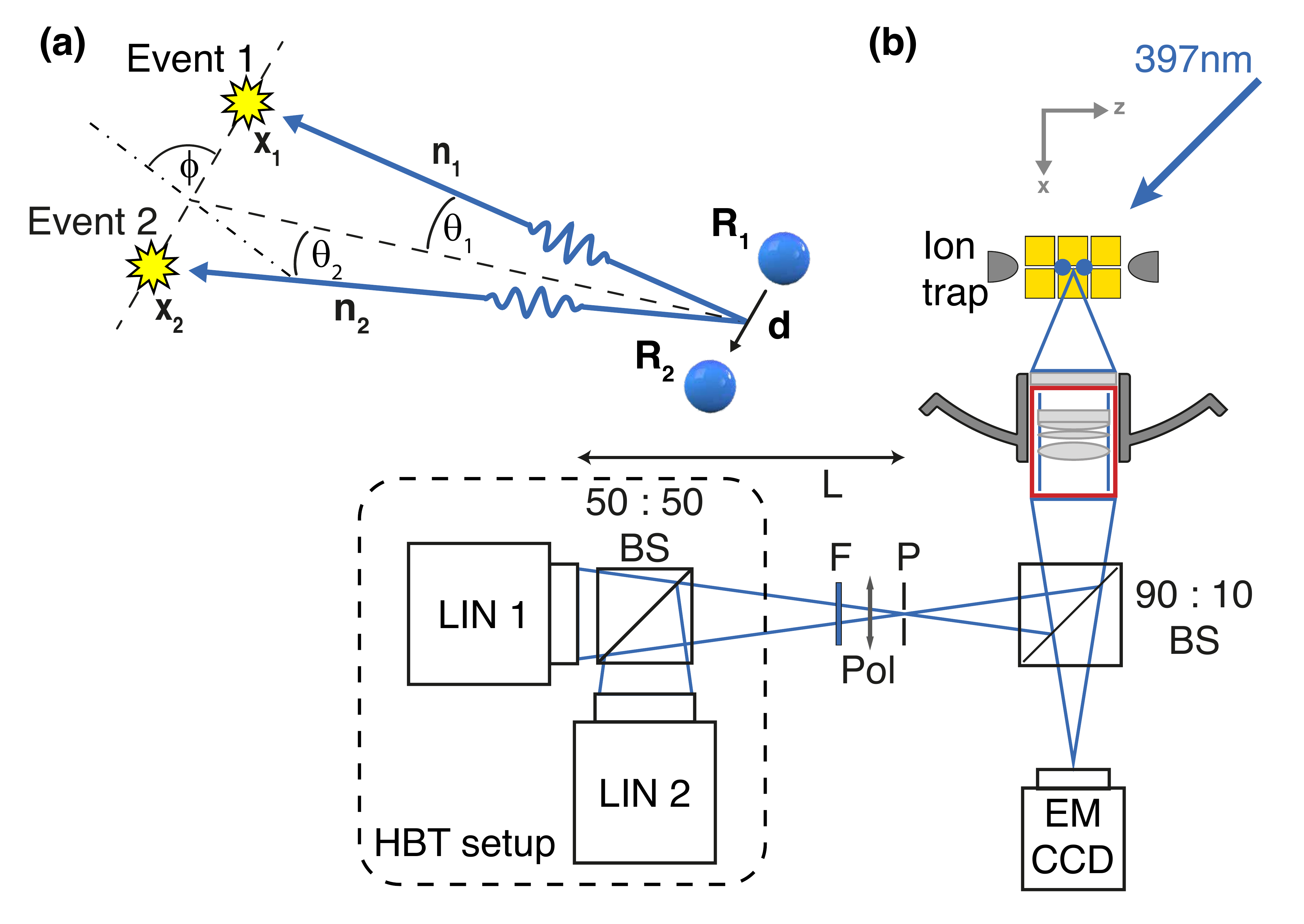}
\caption{(a) Two ions at a distance $\boldsymbol{d}$ are used as model SPE system. 
A pair of fluorescence photons is scattered off the ions into directions $\boldsymbol{n}_1$ and $\boldsymbol{n}_2$ and recorded coincidently at detector positions $\boldsymbol{x}_1$ and $\boldsymbol{x}_2$. 
(b) Experimental setup: two $^{40}\text{Ca}^+$-ions are trapped in a linear Paul trap and cooled by laser light, red detuned to the S$_{1/2}$ - P$_{1/2}$ transition. 
Coincident photon events are recorded by a HBT setup (for details see text).\\\\}
\label{Fig3}
\end{figure}

In our setup we record coincident photon events in the far field on a pixelated ultra-fast camera, see \hyperref[Fig1]{Fig.~1}. The analysis of the cross-correlation signal allows for determining the spatial arrangement of an initially unknown number of SPEs. 
In the case of a single SPE, no cross-correlation signal emerges as only one photon at a time is emitted. For two or more SPEs, various spatial frequencies - governed by the distances between the emitters - are observed in the cross-correlation signal. In principle, one might directly analyze the spatial two-dimensional cross-correlations. However, for situations where the number of recorded two-photon coincidences is low, it is preferable to project the signal onto a single axis. The axis is chosen by maximizing the contrast of the projected one-dimensional cross-correlation signal. This selects a direction which is parallel to the distance vector between the two SPEs, see \hyperref[Fig2]{Fig. 2}. The periodicity of the cross-correlation signal, i.e., the spatial frequency $f_\text{spatial}$, along this axis determines the separation of the two SPEs.

Considering the case of two laser excited immobile SPEs, the coincident two-photon cross-correlation function reads \cite{PhysRevA.64.063801,Wiegner2015}
\begin{equation}
\begin{split}
 G^{(2)}\left( \boldsymbol{x}_1, \boldsymbol{x}_2\right) &=  \langle E^{(-)}(\boldsymbol{x}_1)\,E^{(-)}(\boldsymbol{x}_2)\,E^{(+)}(\boldsymbol{x}_2)\,E^{(+)}(\boldsymbol{x}_1)\rangle  \\
 &  \sim 1+ \cos\left[  \delta(\boldsymbol{x}_1) - \delta(\boldsymbol{x}_2) \right] \;. 
\end{split}
\label{eq1}
\end{equation}
Here, $E^{(+)}(\boldsymbol{x}) =\sum_{l} e^{-i k_L\,\boldsymbol{n}\cdot \boldsymbol{R}_{l}} s_{-}^{(l)} \, $ $ \left[ E^{(-)}(\boldsymbol{x})=E^{(+) \, \dagger}(\boldsymbol{x}) \right]$
denotes the positive [negative] frequency part of the electric field at position $\boldsymbol{x}$, with $s_{-}^{(l)}$ the lowering operator of the $l$th SPE, $l=1, 2$. The term $\delta\left( \boldsymbol{x} \right) = \left( \boldsymbol{k}_L - k_L\boldsymbol{n} \right) \cdot \boldsymbol{d}$ expresses the phase difference accumulated by a photon scattered by SPE1 at $\boldsymbol{R}_{1}$ with respect to a photon scattered by SPE2 at $\boldsymbol{R}_{2}$ and recorded at the same detector pointing in the direction $\boldsymbol{n} = \boldsymbol{x}/|\boldsymbol{x}|$, where $\boldsymbol{d}= \boldsymbol{R}_{2} - \boldsymbol{R}_{1}$ is the distance vector between the two SPE and $\boldsymbol{k}_L$ the wave vector of the driving laser, see \hyperref[Fig3]{Fig. 3(a)}. 

To exemplify our method, we employ trapped ions providing spatially fixed SPEs, see \hyperref[Fig3]{Fig. 3(b)}: two $^{40}\text{Ca}^+$ ions are trapped \cite{PhysRevLett.117.043001} and continuously Doppler-cooled on the  S$_{1/2}$ - P$_{1/2}$ transition using laser light near $396.95\,$nm. In the harmonic potential with trap frequencies $\omega_{(z,R1,R2)} / 2\pi = (0.76,1.275,1.568)\,$MHz we achieve a mean occupation of about $10$ phonons per mode, corresponding to a wave packet size $< 50\,$nm. 
A magnetic field of $0.62\,$mT is applied along the $\boldsymbol{e}_y$-direction to determine the quantization axis of the system. To run the experiment $24/7$ continuously, $10\,\%$ of the fluorescence light is monitored by an auxiliary EMCCD-camera such that in case of ion loss a reloading sequence is automatically launched. 

Under continuous laser excitation near $397\,$nm as well as $866\,$nm for repumping and emptying the metastable D$_{3/2}$ level, photons scattered off the ions are collected by a $f/1.6$ lens at a working distance of $48.5\,$mm and steered into a HBT detection setup consisting of a $50:50$ beam splitter (BS) and two synchronized microchannel plate (MCP) detectors~\footnote{MCP, LinCam by Photonscore} for overcoming the dead time of the MCPs of $600\,$ns. The MCPs provide direct charge readout with $1000 \times 1000$ spatial bins and a timing resolution of $50\,$ps at a maximum count rate of $600\,$kHz per detector, thus combining high spatial \textit{and} temporal resolution. Indistinguishability of the scattered photons with respect to polarization is assured by a polarizing filter (Pol). A pinhole (P) in an intermediate focus and a band pass filter (F) suppress stray light. In the HBT setup we have chosen a coincidence window of $2.5\,$ns, significantly shorter than the lifetime of the excited state of $\tau_{P1/2} = 6.9\,$ns. Under typical operation conditions, we observe a coincidence rate of $\sim 68\,$mHz, while count rates at each detector are $\sim 7\,$kHz. 

After projecting the $1000 \times 1000$ virtual pixels of each MCP onto one dimension, every possible two-photon coincident event $G^{(2)}\left(\boldsymbol{x}_1,\boldsymbol{x}_2 \right)$ is stored in a binned-data structure $G^{(2)}_{i,j}$, encoding $96$ start positions $i$ and $96$ corresponding stop positions $j$. After $756\,$hours of data acquisition each entry of the binned-data structure is filled on average with $20$ events. As outlined above, in order to determine the absolute orientation of the two-ion crystal, we rotate the recorded two-photon coincidences $G^{(2)}\left(\boldsymbol{x}_1,\boldsymbol{x}_2 \right)$ around the angle $\phi$ optimizing for the contrast of the binned-data. This procedure shows a distinct maximum at $\phi = 0.86^{\circ} \pm 0.31 ^{\circ}$, see \hyperref[Fig2]{Fig. 2(a)-(d)}, determining the absolute orientation of the direction of $\boldsymbol{d}$.

To access the distance $d$ between the ions, we extract the spatial frequency $f_\text{spatial} $ from the cosine-fit to the binned-data at optimum contrast, see \hyperref[Fig2]{Fig. 2(a)}. In the far field, and taking into account the magnification of the light collection system $M$, see \hyperref[Fig3]{Fig. 3}, we find for the phase difference as a function of the stop detector position $\delta\left( \Theta_2 \right) = - k_L/\sqrt{2} - k_L \, M d \, \Theta_2$, and thus for the  spatial frequency $f_\text{spatial} =  k_L \, M d$, where $k_L=2\pi/\lambda$ is the wave number of the excitation laser light at 397~nm.  The binned-data $G^{(2)}_{i,j}$ is fit by a cosine for each start position $i = 1, \ldots, 96$, however, we use only the central $i=27, \ldots, 67$ which, due to the circular shape of the MCPs, allows for an unambigous fitting and is comprising $> 52\%$ of the total data. From the fits we determine $f_\text{spatial} = 1490\pm2_{\text{stat.}}\pm8_{\text{syst.}}\,\text{rad}^{-1}$, where the statistical error as a function of the accumulated coincidences follows a power law $N^{(-0.51\pm0.06)}$, with a maximum number of coincidences $N \sim 2 \cdot 10^5$, see \hyperref[Fig2]{Fig. 2(e)}. We account for the systematic uncertainty by measuring the distance between the intermediate image and the MCP detectors to $L=448\pm 1\,$mm, intervening in order to gauge the pixel sizes in angular units $\Theta_2$, see \hyperref[Fig3]{Fig. 3(a)}. In the future, placing the HBT setup at various accurately measured distances $L$ and determining the corresponding $f_\text{spatial}(L)$ would allow for greatly reducing this systematic uncertainty. 

Verifying this outcome by an independent measurement, we derive the ion distance to $6.696\pm 0.006\,\mu$m, using the measured trap frequency of $762.8\pm 1.0\,$kHz of a $^{40}$Ca$^+$ ion along the $z$-axis \cite{james1998quantum}. With a collection lens magnification of $M=14.1\pm 0.1$, this yields a spatial frequency $f_\text{spatial}^\text{th}  = 1494 \pm 11\,\text{rad}^{-1}$. Note, that this independently derived value - within its larger error - fully confirms the outcome based on the $G^{(2)}$ structure analysis outlined above.

\begin{figure}
\includegraphics[width=0.50\textwidth]{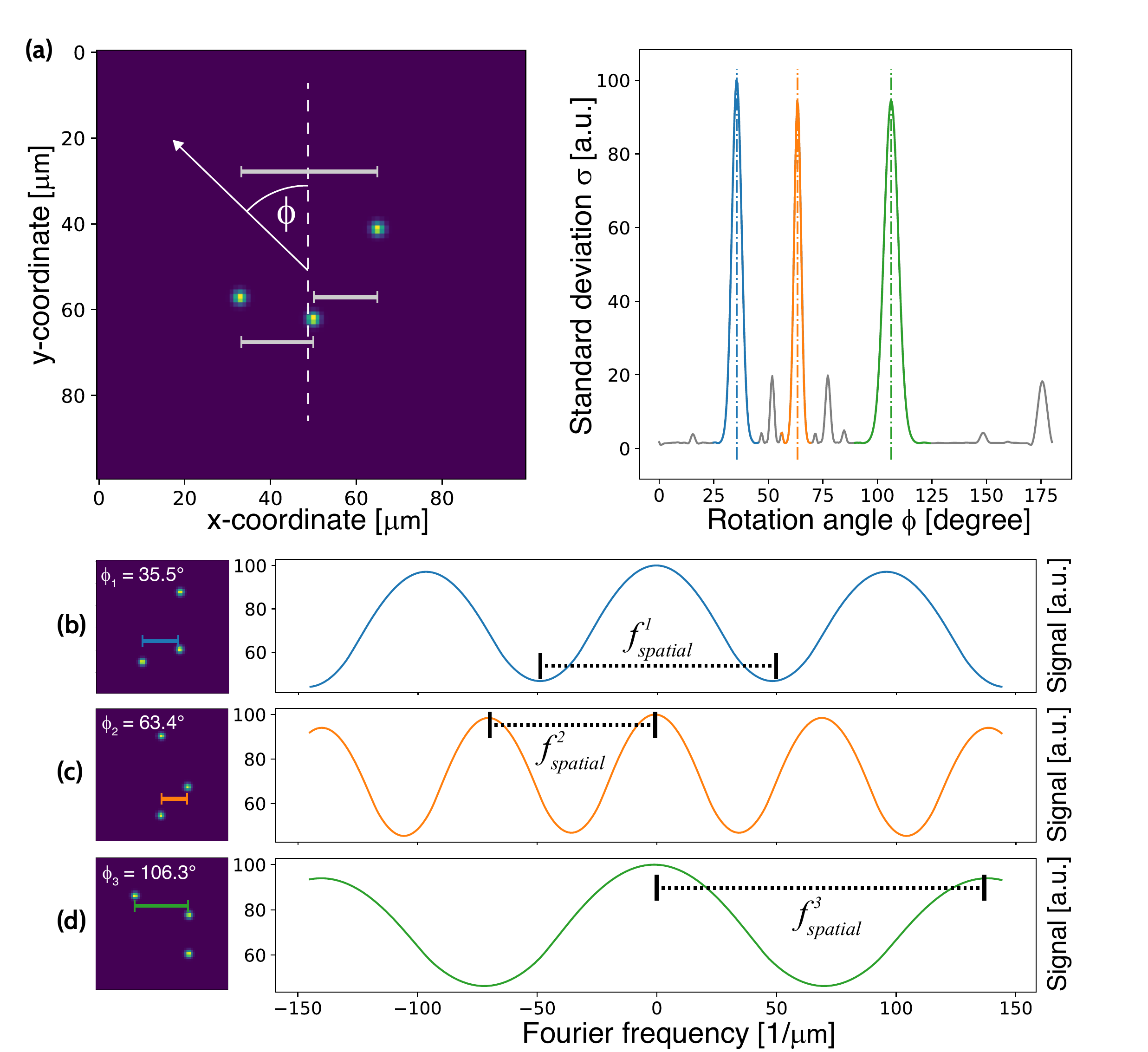}
\caption{Simulating the imaging of a planar three-SPE array:  (a) Analyzing the cross-correlation signal as in \hyperref[Fig2]{Fig. 2} results in different standard deviations of the one-dimensional cross-correlation signal as a function of $\phi$. (b)-(d) For specific $\phi_i$, we find a single spatial frequency $f_\text{spatial}^{(i)}$, $i=1, 2, 3$, suited for extracting the full structural information.}
\label{Fig4}
\end{figure}

For three and more SPEs, several spatial frequencies $f_\text{spatial}^{(i)}$ appear within the SPE array, rendering the determination of the source distribution more challenging. 
Again, under conditions where the coincidence rate is low, a projection of the two-dimensional cross-correlation signal onto one axis is advantageous. For certain rotation angles $\phi^{(i)}$ the standard deviation of the one-dimensional cross-correlation signal displays local maxima, thus allowing for determining the absolute orientation of the SPE, the spatial frequencies and the corresponding distances. In the case of a planar array of three SPEs, we plot the simulated $G^{(2)}_{i,j}$-data for angles $\phi = 35.5^\circ$, $63.4^\circ$ and $106.3^\circ$ where the standard deviation exhibits a local maximum, see \hyperref[Fig4]{Fig. 4}. From the three angles and the corresponding spatial frequencies $f_\text{spatial}^{(1,2,3)}=0.065, 0.090$ and $0.044\,\mu\text{m}^{-1}$, the full structural information of the three-SPE array is accessible.

In the future, we will implement light collection systems with higher numerical aperture to amass more coincidences and achieve faster structure analysis. Besides a reduction in data acquisition time this will enable us to record cross-correlation signals from larger ion structures, or measure higher order $G^{(N)}$ cross-correlation signals \cite{thiel2007quantum,Oppel2012}. As the simulation in \hyperref[Fig4]{Fig. 4} demonstrates, one may employ our new method for the analysis of planar ion structures, e.g., recording the behavior at a structural phase transition between linear and zigzag configurations~\cite{ULM2013}. In the X-ray domain, the advent of more brilliant light sources will facilitate the use of incoherent scattering for extracting structural information, possibly improving on coherent scattering methods used today~\cite{Classen2017}. Our experiments on collective light scattering off ions, where parameters are precisely tunable over a large range, serve here as a model system for paving the way for structure analysis in more complex systems. At the same time, using ion crystals in Paul traps, the array of SPEs can be tailored for understanding the elusive interplay of spatial order, collective properties~\cite{RUI2020} of multi-particle entanglement and cooperative optical response.

\begin{acknowledgments}
SR and JvZ acknowledge support from the Graduate School of Advanced Optical Technologies (SAOT) and the International Max-Planck Research School, Physics of Light, Erlangen. We thank Photonscore GmbH, Brenneckestr. 20, 39118 Magdeburg (https://photonscore.de) for providing the coincidence MCP systems and Andr\'e Weber for the initial calibration and characterization of the MPC systems. JvZ thanks Ralf Palmisano for making contact to Photonscore GmbH. This research is funded by the Deutsche Forschungsgemeinschaft (DFG, German Research Foundation) within the TRR 306 QuCoLiMa (``Quantum Cooperativity of Light and Matter'') -- Project-ID 429529648.
\end{acknowledgments}
\bibliography{lit1}

\end{document}